\documentclass[a4,twocolumn]{article}
\usepackage{graphicx}

\newcommand{\be}{\begin{equation}}
\newcommand{\ee}{\end{equation}}

\title{Characterization of Multispecies Living Ecosystems \\
With Cellular Automata}

\author{Xin-She Yang \\
{\small Faculty of Engineering, University of Wales Swansea,
Singleton Park, Swansea SA2 8PP,  UK } }

\date{}
\begin{document}
\maketitle
\begin{abstract}
A multispecies artificial ecosystem is formulated using cellular
automata with species interactions and food chain hierarchy. The
constructed finite state automaton can simulate the complexity
and self-organized characteristics of the evolving multispecies
living ecosystems. Numerical experiments show that a small
perturbation or extinction event may affect many other species in
the ecosystem in an avalanche manner. Both the avalanches and the extinction
arising from these changes follow a power law,
reflecting that the multispecies living ecosytems have the
characteristics of self-organized criticality. \\
\end{abstract}

\noindent {\bf Citation detail:} X. S. Yang, Characterization of multispecies living 
ecosystems with cellular automata, in: {\it Artificial Life VIII} (Eds. Standish, Abbass, Bedau),
MIT Press, pp. 138-141 (2002). 

\section{Introduction}

The modelling of artificial life is very important to the
understanding of the origin of life and its evolution, thus the
studies using computer simulations are of both scientific
importance and practical applications.  Since Langton (1986)
relauched the field of artificial life,  the possibly
universal aspects of living systems have been investigated
by exploring the artificial
chemistry acting on artificial molecules in terms of rule-based
cellular automata (CA). Langton (1986) carried out numerous CA
simulations with a parameterization scheme
allowing the relationship of Wolfram's
classes cataloged the rules that generates different classes of
dynamics. Since then there have been many extensive studies
about artificial life. However, the understanding and studies of
artificial life is still at very early stage.

Artificial life simulations using cellular automata have received
much of interests in the community of artificial life and
evolutionary computing. Conway's game of life has popularized this
area of research. However, most of the existing simulations of CA
is about the evolutionary and classification of one single
species, and this has demonstrated the complexity and richness of
simple rule-based systems. Interestingly, multispecies models
have begun to attract much attention (Sole and Manrubia, 1996;
Amaral and Meyer,1999). In order to study the ecological effect
and the interaction among different species, a multispecies system
of artificial life should be explored in detail. It can be
expected that the behavior
and characteristics of multispecies living system could be very
different from that of single species cellular automata.

The aim of this paper is to present some new results using a
multspecies system of artificial life with cellular automata. By
numerical experiments, we will find how the interactions among
different species affect the evolution behavior and self-organization
as well as species extinction events. This paper is
orgainised as follows. In the next sections, we describe how to
construct the cellular automata for the multispecies living systems,
we then implement the systems and give some numerical simulations.
Based on the numerical simulations, the complexity and entropy of
the evolving living systems are measured, and the self-organized
criticality is tested. The possible reason of extinction behavior
of the living systems is also explored. The implications of the
multispecies interactions will be also discussed.

\section{Rule-Based Cellular Automata for Multi-Species Living Systems}

Cellular Automata (CA) is a finite-state machine on a regular
lattice. The input to the machine is the states of all machines in
its neighborhood, the change of its state is based on the rules or
transition functions. The states of all machines in the lattice
are updated synchronously in discrete time steps. Each cell
evolves according to the same rules which depend only on the state
of the cell and a finite number of neighboring cells, and the
neighborhood relation is local and uniform. One classic example
is Conway's game of life where one takes a neighborhood consisting
of the nearest 8 cells to a cell on a 2-D cellular automata.
Its transition function for the local automaton for
two states (1 and 0, or alive and not alive) is as follows:
If 2 or 3 neighbors of a cell are alive (or 1) and it
is alive at present, then it is alive at next state;
If 3 neighbors of a cell are alive and it is currently not alive,
its next state is alive; the next state is not alive for
all the other cases. Even with these simple rules,
an universal computational machine can be constructed on an
infinite 2-D grid, which is capable of emulating the computing
power of any Turing machine or digital computer.

The present model is an extension of the combined
version of the Sole-Manrubia model (1996) and
Amaral-Meyer food chain model (1999). The next state
for a cell is determined by the transition function in
the general form
\be
C_i(t+1)=H [\sum_{j} G_{ij} C_i(t)-\theta],
\ee
where H is the heaviside step function, and $\theta$ is the
threshold. $G_{ij}$ of the influence of cell $j$ on cell $i$,
$G_{ij} >0$ or $<0$ if $j$ is the food  or predator for cell $i$.
The summation is over the nearest neighborhood.
The multispecies cellular automata works in the following way:
\begin{enumerate}
\item The CA consists of $M$ interacting species,
      when $M=1$ it degenerates into the classic Conway's
       game of life and obeys the same simple rules.

\item Different sepcies interacts with each other in a way of food chain
      hierarchy.  Each species is label as a hierarchy level
      and for simplicity, we use the level $i$ for species $i$
      with species $i-1$ as the pray  or food for species $i$ and
      species $i+1$ as its predator.

\item In the nearest 8 cells of to a cell of species $i$,
      if the number of predators is greater than the
      number of prays, and the cell is currently alive, then
      its next state is alive;  If the number of the prays
      is more than the number of predators, then its next state
      is alive whether its present state is alive or not;
      If the predators are in the same number as the
      pray, the present state  does not change.

\item If there is no predator or pray in the 8 neighborhood cells,
      then the transition function is the same as the single species
      Conway's cellular automata.
\end{enumerate}

In the numerical simulations, the 2-D lattice is randomly initialized with
the highest population of the lowest level of species $i=1$. For a regular
100 x 100 lattice with N=256 species initialized this way, we can
investigate the complexity, self-organization,
species interaction and avalanche events.

To simulate the effect of species interaction and extinction,
a small probability of extinction rate is also introduced into
some species, the influence of the  extinction of one species on the
other species in the food chain model can be studied in some detail.
In addition, by introducing some extra population in one species,
one can control some other species in the food chain model
and give important implications on the ecological effect in reality
such as the rat control. A number of numerical experiments have also
been investigated. Some trend and self-organized  criticality is analyzed
from the simulated results.

\section{Numerical Experiments}

By using the cellular automata described in the above section, we
can simulate the behavior of multspecies living systems with
multispecies interactions in a food chain hierarchy. In the rest
of this paper, we present some results from a huge number of
the numerical simulations and parametric studies. A typical initial
configuration is shown in Figure 1.

\begin{figure}[t]
\centerline{\includegraphics[width=3in]{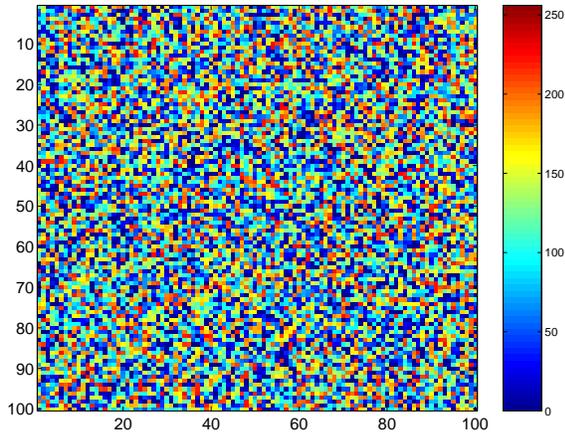}}
\vskip 0.25cm
\caption{A random configuration of multispecies cellular automata}
\label{fig1}
\end{figure}

\subsection{Variation in Complexity and Entropy}

For a population size $N=100 \times 100$ with $M=256$, the
complexity of the cellular automata can be measured by its entropy $S$
\be
S=-\sum_{i} p_i \log p_i,
\ee
where $p_j$ is the probability of find the species $i$ in the total
population (Adami,1998). For a finite population size,
$p_i$ can be approximated by the fraction of samples $n_i$
\be
p_i = \frac{n_i}{N},
\ee
so that
\be
S=-\sum_{i} p_{i} \log p_{i}.
\ee
The variation of complexity of the 256-species cellular automata
is shown in Figure 2.

\begin{figure}[t]
\centerline{\includegraphics[width=3in]{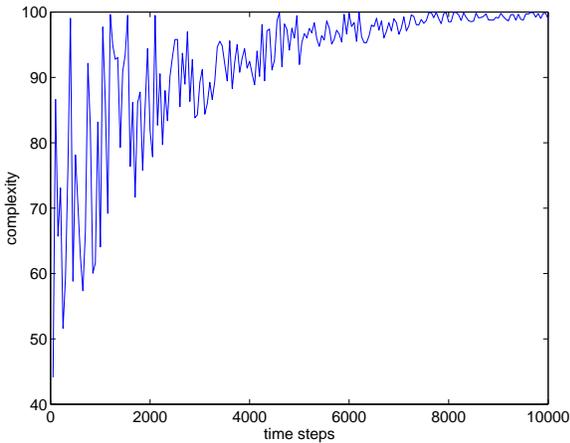}}
\vskip 0.25cm
\caption{Variations of complexity versus time steps}
\label{fig2}
\end{figure}

It is clearly seen that the complexity varies significantly at the early
stage of the evolution process, then it gradually relaxes to the
equilibrium  at long time, indicating that the living systems
is in a quasi-steady state among different species.

\subsection{Self-Organized Criticality}

Based on the pioneering work by Adami (1995) on the self-organized
criticality in living systems using the {\tt Tierra} experiments
and the subsequent work by Newman et al (1997), there is a good
reason to believe that self-organized criticality exists in the
artificial living systems.
In order to test whether the self-organized criticality exists
in the multispecies living ecosystems, we introduce a small perturbation
to ecosystem and measure the population changes
affected by the perturbation or avalanche in population and species,
this is because the ecosystem can form between all the competing
and yet interconnected species. The change in response to any
perturbation will enable this living to relax to a new equilibrium
or the self-organized state. By tracing the 15000 avalanches obtained
from numerical simulations on 100 x 100 lattice grid, the results
are shown in the log-log plot in Figure 3.

\begin{figure}[t]
\centerline{\includegraphics[width=3in]{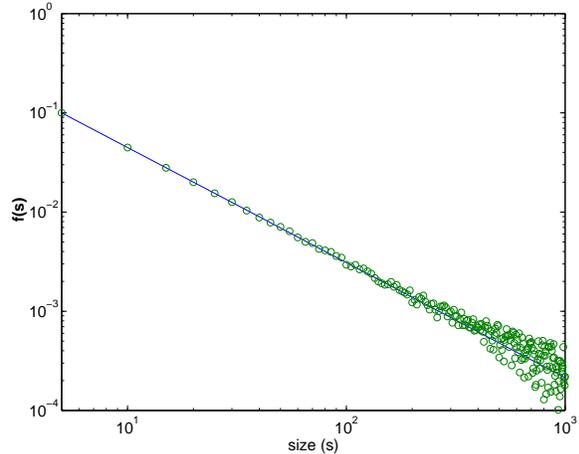}}
\vskip 0.25cm
\caption{Self-organized criticality in multispecies living systems}
\label{fig3}
\end{figure}

This figure clearly show that the avalanches size distribution
follows a power-law with the exponent $\gamma=2.26 \pm 0.04$.
This implies that the multispecies living ecosystem has the
characteristics of self-organized criticality. This is consistent
with the earlier work by Adami (1995). Our present work implies that
the self-organized criticality could be universal in living systems.

\subsection{Extinction}

In the above numerical simulations, the changes or
redistribution of populations of different species have been studied.
There may not necessarily involve the extinction of some species.
In some case, for example, when the initial population size for
some species is very small and its predators are strong, the whole species
could become extinct, and the extinction of one species
may influence other species in the food chain. To simulate this
effect, we introduce an extinction probability $p_e=0.01$ for a species
at some level, say $i=16$, in the food chain hierarchy. The extinction
obtained from the numerical simulations is shown in Figure 4.

\begin{figure}
\centerline{\includegraphics[width=3in]{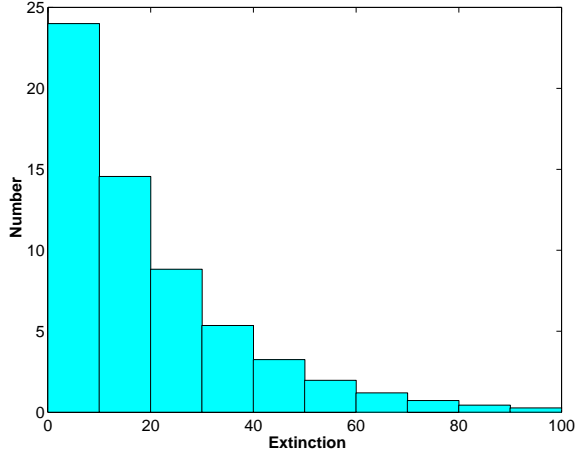}}
\vskip 0.25cm
\caption{ Extinction intensities at different time stage }
\label{fig4}
\end{figure}

The extinction intensities at different time stages (100 time steps as a stage)
follow a power-law with an exponent $\gamma=2.05 \pm 0.1$, which
implies that extinction is a self-organized phenomenon. This is consistent
with the fitting data $\gamma=1.7 \pm 0.3$ from the fossil record
(Newman and Eble, 1999), and the extinction is a result of
external change and species competition.

\section{Discussion}

The multispecies ecosystems of artificial life have been formulated in
terms of 2-D cellular automata of the combined Sole-Manrubia
type together with the Amaral-Meyer food chain model. By using the proper
interactions between different species in the food chain hierarchy
and the transition function, the finite state automaton can simulate
the complexity of the evolving multispecies living ecosystem.
Numerical experiments show that the complexity measured by the
system entropy fluctuates at early stage of evolution, then
it evolves to the self-organized or dynamic equilibrium sate under
certain conditions. By introducing the extra fraction of population
or a small perturbation, the species population may be affected at
different scales. The avalanches arising from these changes obey
a power-law with the exponent $\gamma=2.26$,
reflecting that the multispecies living ecosystem has the
characteristics of self-organized criticality.

One the other hand, in the case of one species in the margin of
extinction, it can affect many species in the multispecies
living systems and can lead some kinds of mass extinction. Numerical
simulations also show that the extinction intensities follow a
power-law form with the exponent $\gamma=2.05$,
thus indicating the species interaction and competition
is the mechanism for the extinction under the action of external
changes. Although the present modelling provides a lot of
features about the complexity of the artificial living systems,
much more work is clearly needed to investigate the system
behavior such as how the different extinction mechanisms
and transition rules affect the evolution of the multispecies
living ecosystems.

\section*{Reference}

\begin{description}

\item Adami, C (1995).
      Self-organized criticality in living systems,
     {\it Phys. Lett. A}, 203:23.

\item Adami, C (1998).
      Introduction to artificial life, Springer-Verlag, New York.

\item Amaral, L. A. N. and Meyer, M. (1999). Environmental changes,
      coextinction, and patterns in the fossil record.
      {\it Phys. Rev. Lett}. 82:652.

\item Flake, G A (1998). The computational beauty of nature, MIT press.

\item Langton (1986). Studying artificial life with cellular automata,
{\it Physica D}, 22:120.

\item Newman, M. E. J. and Eble, G. J. (1999). Power spectra
     of extinction in the fossil record. {\it Proc. R. Soc. London},
      B266:1267.

\item Newman, M E J, S M Fraser, K Sheppen and W A Tozier, (1997).
Comment on ``Self-organized criticality in living systems'' by C Adami,
{\it Phys. Lett. A}, 228:201.

\item Sole, R. V. and Manrubia, S. C. (1996). Extinction and
      self-organized criticality in a model of large-scale evolution
      {\it Phys. Rev. E}, 54:R42.

\end{description}

\end{document}